\documentclass[12pt,a4paper]{article}
\usepackage[english]{babel}
\usepackage[T1]{fontenc}
\usepackage[utf8]{inputenc}
\usepackage{ragged2e} 
\usepackage[bookmarks=false,linkbordercolor={1 0 0},citebordercolor={0 1 0}]{hyperref}
\usepackage{xcolor}
\hypersetup{
    colorlinks=True,
    urlcolor = black,
    citecolor = red,
    linkcolor=red,
    }
\usepackage[left=2.00cm, right=2.00cm, top=2.00cm, bottom=2.00cm]{geometry}
\usepackage{authblk}
\usepackage{graphicx, caption}
\title{\Large\textbf{Fine-grained All-fiber Nonlocal Dispersion Compensation in the Telecommunications O-Band}}
\author[1,2\thanks{\color{blue}{ruiming.chua@tii.ae}},3]{RuiMing Chua}
\author[1,2]{James A. Grieve}
\author[1,3]{Alexander Ling}
\affil[1]{Centre for Quantum Technologies, 3 Science Drive 2, National University of Singapore, 117543 Singapore}
\affil[2]{Quantum Research Centre, Technology Innovation Institute, Abu Dhabi, UAE}
\affil[3]{Department of Physics, National University of Singapore, Blk S12, 2 Science Drive 3, 117551 Singapore}

\date{\today}
\begin{document}
\justifying
\sloppy
\maketitle
\begin{abstract}

Nonlocal dispersion compensation between broadband photon pairs propagated over fiber corresponding to the ITU-T G.652D telecommunications standard was studied via fine-grained measurements of the temporal correlation between them. We demonstrated near-ideal levels of nonlocal dispersion compensation by adjusting the propagation distance of the photon pairs to preserve photon timing correlations close to the effective instrument resolution of our detection apparatus (41.0$\pm$0.1ps). Experimental data indicates that this degree of compensation can be achieved with relatively large fiber increments (1km), compatible with real-world deployment. Ultimately, photon timing correlations were preserved down to 51ps$\pm$21ps over two multi-segmented 10km spans of deployed metropolitan fiber.\break
\end{abstract}

\section{Introduction}

Correlated photon pairs produced via Spontaneous Parametric Down Conversion (SPDC) are an important resource for various quantum technologies such as entanglement-based quantum key distribution protocols and entanglement-based clock synchronization\cite{Lee:16,Xu2019Energy,Shi2020stable,Lee2019symmetrical}. These photon pairs possess a high degree of temporal correlation (typically 10fs to 100fs) that is essential for such quantum technologies\cite{Lee:16,Xu2019Energy,Shi2020stable,Lee2019symmetrical,Kevin2011Observations}. However, the SPDC process is inherently broadband which enables chromatic dispersion to easily obscure these correlations\cite{Yang2008Effects}. For fiber-based quantum technologies utilizing SPDC, this is a significant problem which is often mitigated by spectral filtering or compensation of chromatic dispersion\cite{Xu2019Energy,Wengerowsky2019Entanglement}.\break

\begin{figure}[hbt!]
\centering\includegraphics[width=8cm]{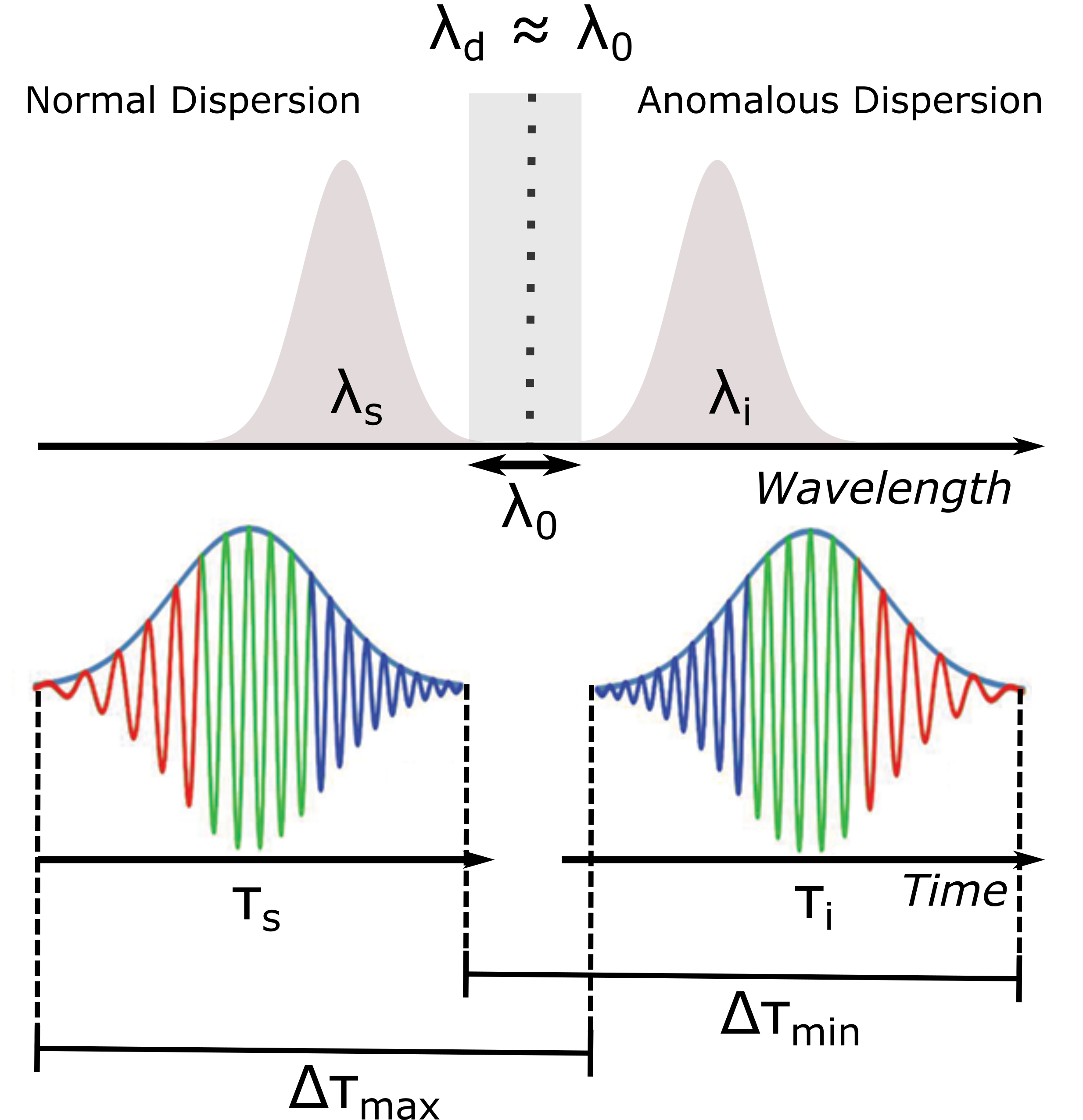}
\captionsetup{width=.8\linewidth}
\caption{A light pulse experiencing anomalous dispersion has higher energy (shorter wavelength) components travelling faster than lower energy (longer wavelength) components while the opposite holds for normal dispersion\cite{Mitschke2016Fiber}. In the SPDC process, high/low energy components of the signal spectrum are correlated with low/high energy components of the idler spectrum. As a result of this energy anti-correlation, fast/slow-propagating components of the signal spectrum becomes matched with fast/slow-propagating components of the idler spectrum, minimising the discrepancy between the minimum and maximum possible delays $\Delta$$\tau$\textsubscript{min} and $\Delta$$\tau$\textsubscript{max} incurred between the correlated photon pairs which in turn relates to the spread in observed timing correlations\cite{Grieve2019Characterizing}. As a result, the chromatic dispersion experienced by the signal and idler spectra will cancel each other out, preserving the degree of coincidence between the correlated photon pairs.\parfillskip0cm}
\label{Figure 1}
\end{figure}

In 1992, Franson showed that SPDC photon pairs, entangled in the time-energy basis, experience nonlocal compensation of chromatic dispersion when the complementary photons propagate through media with opposite signs in dispersion coefficients (Figure \ref{Figure 1})\cite{Franson1992Nonlocal}. This phenomenon may be quantitatively understood in terms of its effect on the timing distribution between the correlated photon pairs:\break

\begin{equation}
{\sigma^2} = {2\sigma_0^2} + \frac{({\beta_1}x_1 + {\beta_2}x_2)^2}{2\sigma_0^2},
\end{equation}
\label{Equation 1}

Where $\sigma$ is the width of the timing distribution, $\sigma_0$ is the intrinsic spread of the correlated photon pairs, $\beta_1$ \& $\beta_2$ are the dispersion coefficients and $x_1$ \& $x_2$ are the propagation distances. When the coefficients or the propagation distances are negligibly small, the width of the timing distribution reduces to $2\sigma_0^2$. In the opposite limit where the coefficients or the propagation distances are large, the width of the timing distribution may also be minimized if $\beta_1$$x_1$ \& $\beta_2$$x_2$ have the same magnitude but opposing signs\cite{Franson1992Nonlocal,Grieve2019Characterizing}.\parfillskip0cm

The concept of nonlocal dispersion compensation was later expanded upon to reemphasize the phenomenon as an unambiguous feature of quantum entanglement in the presence of dispersive transmission\cite{Wasak2010Entanglement}. Experimentally, it has been investigated with the use of external dispersive elements, such as prisms, and single-mode optical fibers (specified by the International Telecommunications Union standards ITU-T G.652/7) exhibiting opposite dispersion coefficients typically near the 1310nm “O-band” of fiber-optic communication\cite{Baek:09,Fiberlabs,ITU-TG652,ITU-TG657}. Earlier experiments with the use of single-mode optical fibers utilized a tunable source of SPDC photons to produce wavelengths that would experience dispersion compensation when propagated through two continuous spans of deployed telecommunications fiber with lengths of up to 9.3km\cite{BRENDEL1998Measurement}. Subsequently, it was shown that broadband SPDC photons can also exhibit nonlocal dispersion compensation without deliberately filtering the source\cite{Grieve2019Characterizing}, thereby simplifying and increasing the throughput of the entire system.\break

The focus of this study was to examine the practical limit of this scheme. Dispersion and nonlocal dispersion compensation experienced by a nondegenerate broadband ($\approx$30nm FWHM) SPDC source over G.652D telecommunication fiber was measured with high resolution. We show that this scheme can preserve the temporal correlation of our photons close to the limit imposed by our instrument resolution of 41ps. Furthermore, this scheme is easy to implement, simply achieved with coarse adjustments to the propagation distances.{\parfillskip0cm}

\section{Experiment}

\subsection{Set-Up \& Method\raggedright}

\begin{figure}[hbt!]
\centering\includegraphics[width=10.5cm]{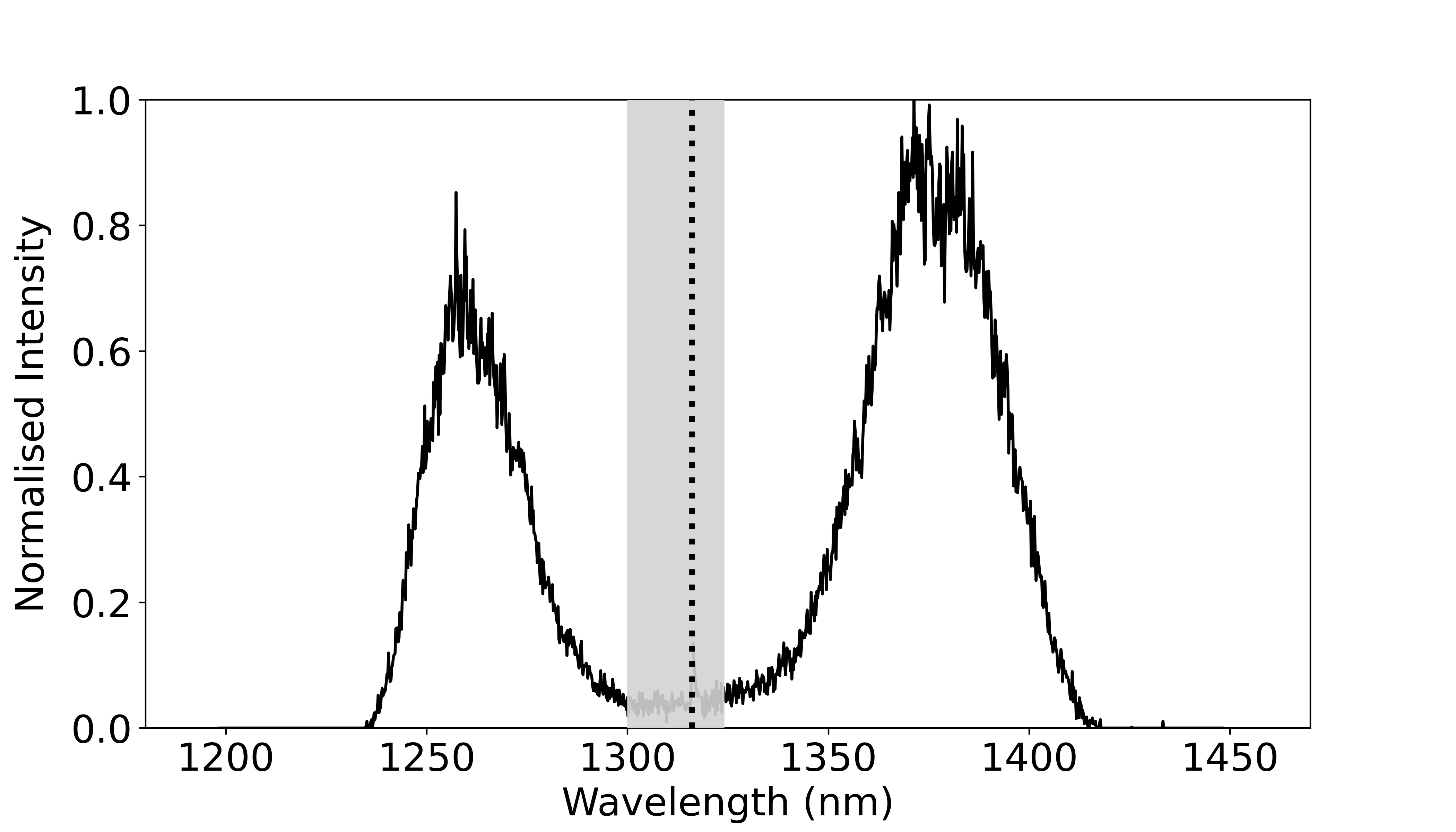}
\captionsetup{width=.8\linewidth}
\caption{Spectrum of our photon pair source. The spectrum of the signal photons, the higher energy photon of each pair (left of vertical line), spectrum of the idler photons, the lower energy photon of each pair (right of vertical line), degenerate wavelength (vertical line) and the range of zero dispersion wavelengths (shaded) of ITU-T G.652/7 single-mode optical fibers are indicated within the spectrum.\parfillskip0cm}
\label{Figure 2}
\end{figure}

\begin{figure}[hbt!]
\centering\includegraphics[width=10.0cm]{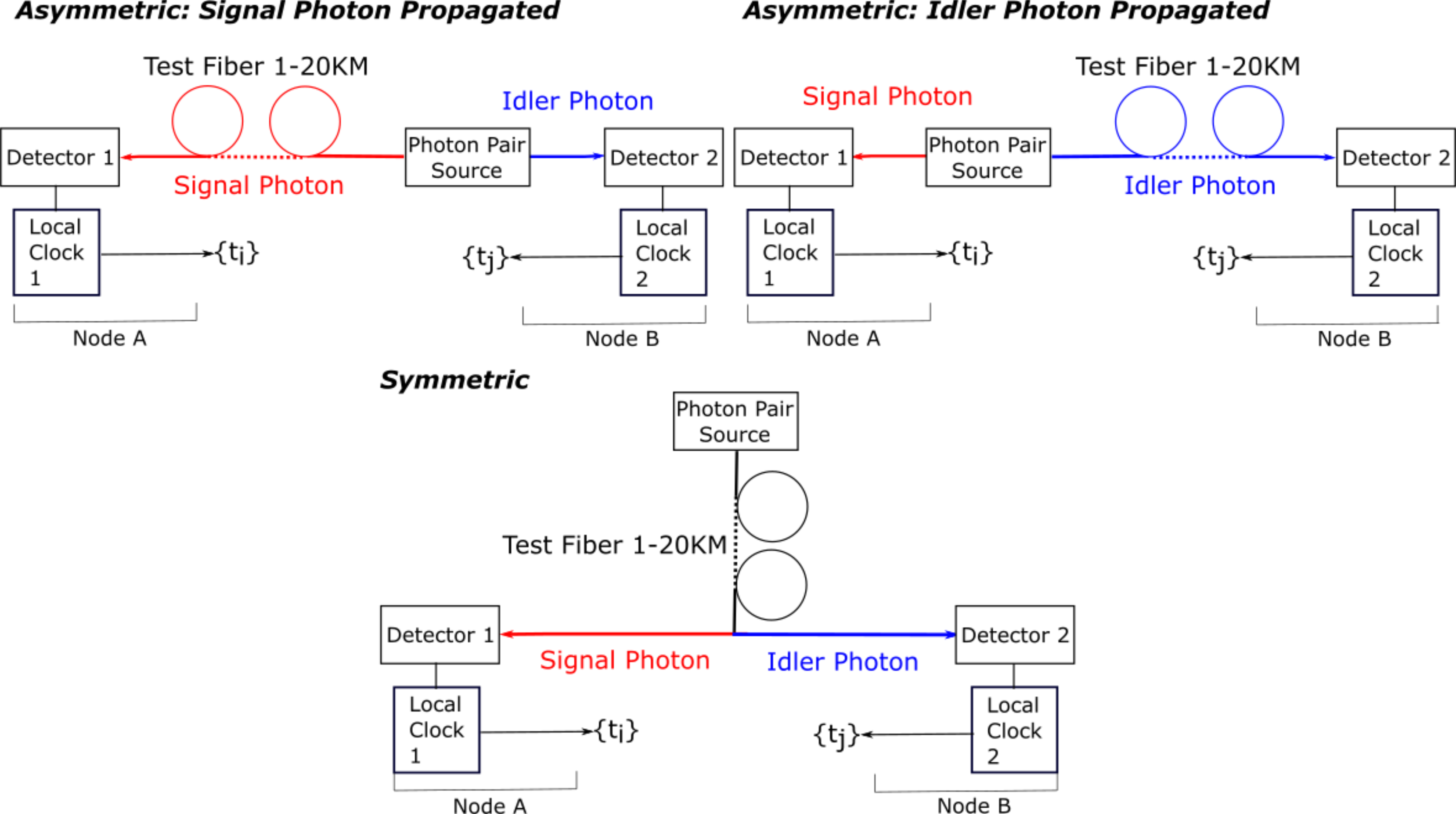}
\captionsetup{width=.8\linewidth}
\caption{Various configurations of our experimental set-up. Asymmetric configurations enable C($\tau$) measurements to be performed when only the signal or idler photons were dispersed while the other was sent directly into the detectors. Asymmetric configurations capture the effect of dispersion on the temporal correlation of the broadband correlated photon pairs. Symmetric configurations enable C($\tau$) measurements to be performed when both photons were transmitted through the same fiber before being separately detected. The results of the symmetric configuration, when juxtaposed with the results of the asymmetric configuration, captures the effect of nonlocal dispersion compensation on the temporal correlation of these photon pairs.\parfillskip0cm}
\label{Figure 3}
\end{figure}

Our study utilized a source based on Type-0 SPDC in a periodically poled crystal of potassium titanyl phosphate (PPKTP, Raicol). The crystal was pumped by a grating stabilized laser diode at 658nm (Ondax) to produce photon pairs in the region of 1316nm\textemdash within the region of possible zero dispersion wavelengths of the G.652 single-mode fibers\cite{ITU-TG652}. The crystal was tuned by temperature to produce a highly nondegenerate spectrum which enabled the photon pairs to firstly, be efficiently separated using a wavelength division demultiplexer, and secondly, experience opposite dispersion coefficients\textemdash the prerequisite for nonlocal dispersion compensation (see Figure \ref{Figure 2}). The wavelength division demultiplexer (WDM, FONT Canada) was centered at 1322nm. Variable lengths of optical fiber were inserted before or after the WDM based on the type of measurement taken to characterize the extent of nonlocal dispersion compensation (Figure \ref{Figure 3}). All the optical fiber used for the experiment was SMF-28e single-mode optical fiber (Corning) conforming to the G.652D specifications\cite{CorningDatasheet}. The photons were detected by superconducting nanowire detectors (SNSPDs) optimized for the O-band, each with a timing jitter of 18ps (Single Quantum Eos)\cite{SingleQuantum}. Events were recorded with a per-channel resolution of 9ps (Time Tagger Ultra, Swabian Instruments)\cite{Swabian}. The system jitter increases with count rate; hence, the single photon count rate was kept close to but not exceeding 200k counts/s to keep the combined measurement jitter of the system below 41.0$\pm$0.1ps for every measurement.

All our fiber segments were characterized for their dispersion properties, including the group delay curve, zero dispersion wavelength, and dispersion profile. We employed a technique based on fitting a group delay curve to measurements of the time-of-flight for a range of wavelengths selected using a monochromator\cite{ANDERSON2004359}. The fitting equation was provided by the three-term Sellmeier equation for G.652/7 series optical fibers that results in the group delay curve (Equation \ref{Equation 2}) with the resulting dispersion curve (Equation \ref{Equation 3}) given by\cite{ANDERSON2004359}:

\begin{equation}
{A\lambda^2 + B\lambda^{-2} + C,}
\label{Equation 2}
\end{equation}
\begin{equation}
{2A\lambda-2B\lambda^{-3},}
\label{Equation 3}
\end{equation}
\break
where $\lambda$ is the wavelength and A, B, and C are empirically determined coefficients. Characterization of nonlocal dispersion compensation relied primarily on the measurement of the temporal correlation between the broadband correlated photon pairs. This is achieved by performing a cross correlation between the detection times of the signal and idler photons. The detrimental effect of dispersion manifests itself with a reduction in the rate of coincidence between the photon pairs, accompanied with the increase in the width of the cross correlation function’s (C($\tau$)) peak (degree of coincidence). Conversely, any preservation in the peak’s width over the same length of fiber between the photon pairs indicates the presence of nonlocal dispersion compensation. C($\tau$) was measured by the time tagger with a chosen bin width of 1ps. The system jitter provided the lower bound to the width of the C($\tau$) peak. Conversely, width of the C($\tau$) peak when only one of the paired photons produced in the SPDC process was sent through the fiber corresponded to measurements of dispersion. In the characterization of nonlocal dispersion compensation, variations in fiber properties were minimized by only using fiber patches cut from a single “parent” spool. Measurements were performed in intervals of 1km for up to 20km with the lengths obtained by connecting these fiber patches in series. For our field segment of the experiment, we utilized two multi-segmented 10km deployed fibers connected to the lab in a “loopback” configuration.\parfillskip0cm

\subsection{Characterizing Nonlocal Dispersion Compensation\raggedright}

\begin{figure}[hbt!]
\captionsetup{width=.8\linewidth}
\centering\includegraphics[width=10.5cm]{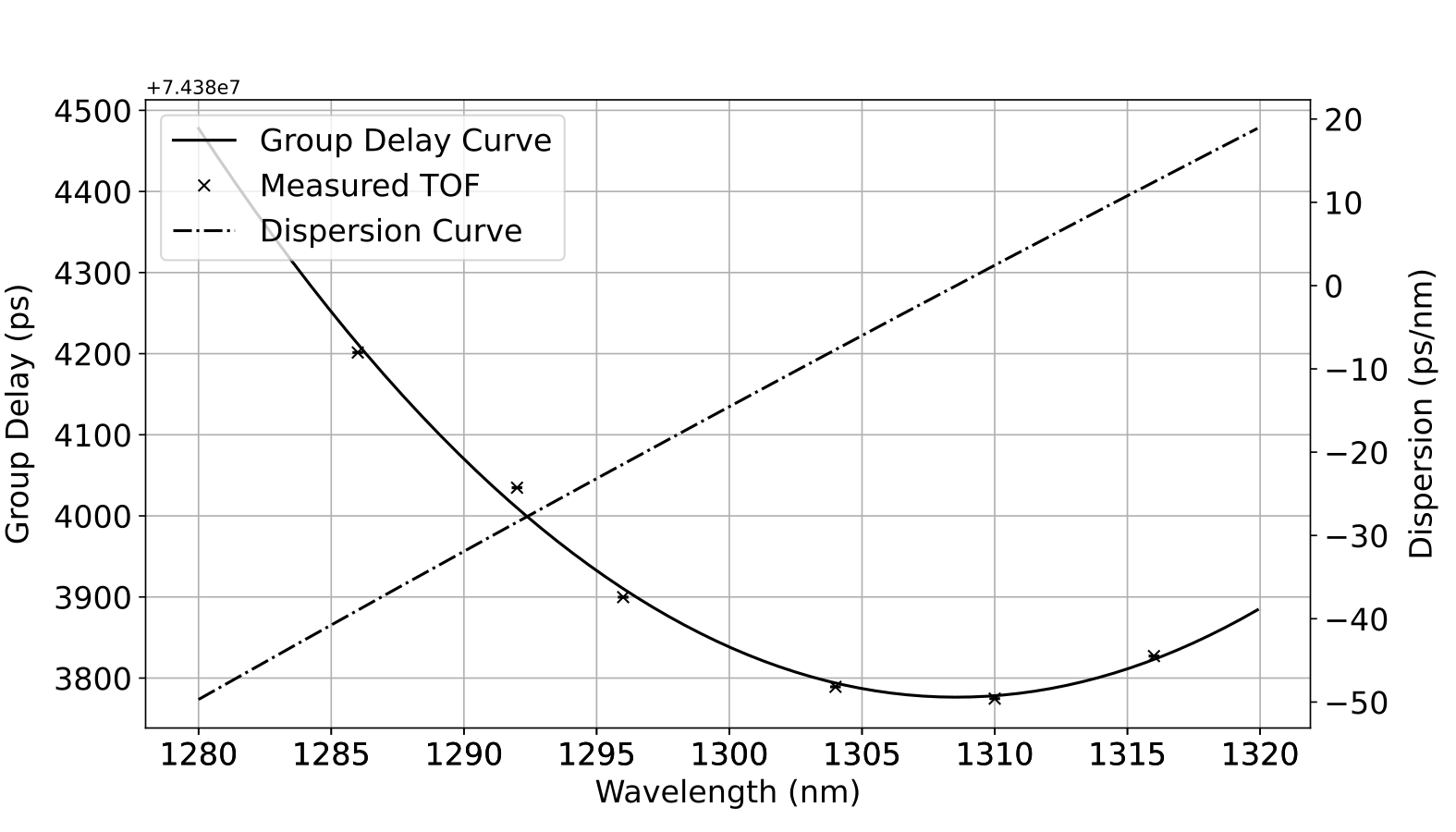}

\caption{An example of the characterization data. Group delay and dispersion curves for a multi-segmented 15km long piece of optical fiber used in the experiment (error bars negligibly small). A group delay curve based on the three-term Sellmeier equation was fitted into the time-of-flights (TOF) of photons at six selected wavelengths. The optical fiber was found to have a group delay curve with coefficients of A=2.10$\times$$10^{-1}$, B=6.16$\times$$10^{11}$, C=7.34$\times$$10^7$ respectively corresponding to a zero dispersion wavelength of 1308.5$\pm$0.2nm. The result falls within the range expected of G.652/7 single-mode optical fibers.}
\label{Figure 4}
\end{figure}

\begin{figure}[hbt!]
\captionsetup{width=.8\linewidth}
\centering\includegraphics[width=10.5cm]{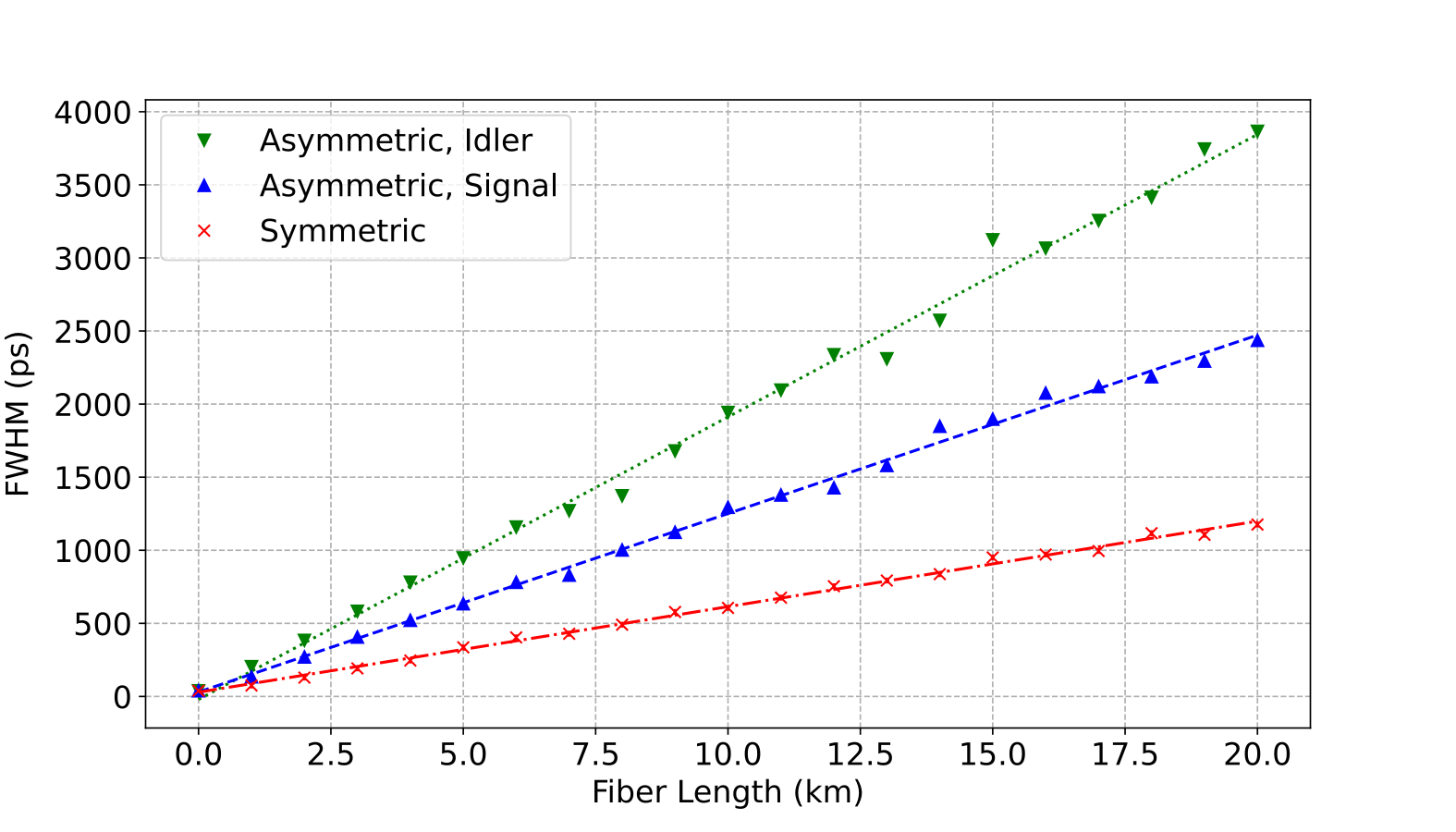}

\caption{FWHM of correlation histogram peaks corresponding to propagation through various lengths of lab fiber (error bars negligibly small). The idler spectrum exhibits dispersion resulting in a broadening of C($\tau$) of 193.3$\pm$3.3ps/km. The signal spectrum exhibits dispersion resulting in a broadening of C($\tau$) of 122.1$\pm$1.6ps/km. Collectively (symmetric case), they exhibit an effective dispersion resulting in in a broadening of C($\tau$) of 58.50$\pm$0.78ps/km.}
\label{Figure 5}
\end{figure}

\newpage

In our experimental characterization of nonlocal dispersion compensation, all our fiber segments were found to have a zero dispersion wavelength of less than 1316nm (mostly $\approx$1310nm), recalling that we utilized fiber segment cut from a single “parent” spool. It can be directly inferred that, coupled to our photon source, our signal photons would be less dispersed than our idler photons (compare Figures \ref{Figure 2} \& \ref{Figure 4}). The outcome of the cross correlation functions conformed to our expected trends (see Figure  \ref{Figure 5}). Nonlocal dispersion compensation was evidenced from the preservation of the degree of coincidence in the symmetric configuration when compared to either of the other cases (compare Figures \ref{Figure 3} \& \ref{Figure 5}). \break

\subsection{Near-Ideal Compensation\raggedright}

\begin{figure}[hbt!]
\captionsetup{width=.8\linewidth}
\centering\includegraphics[width=15cm]{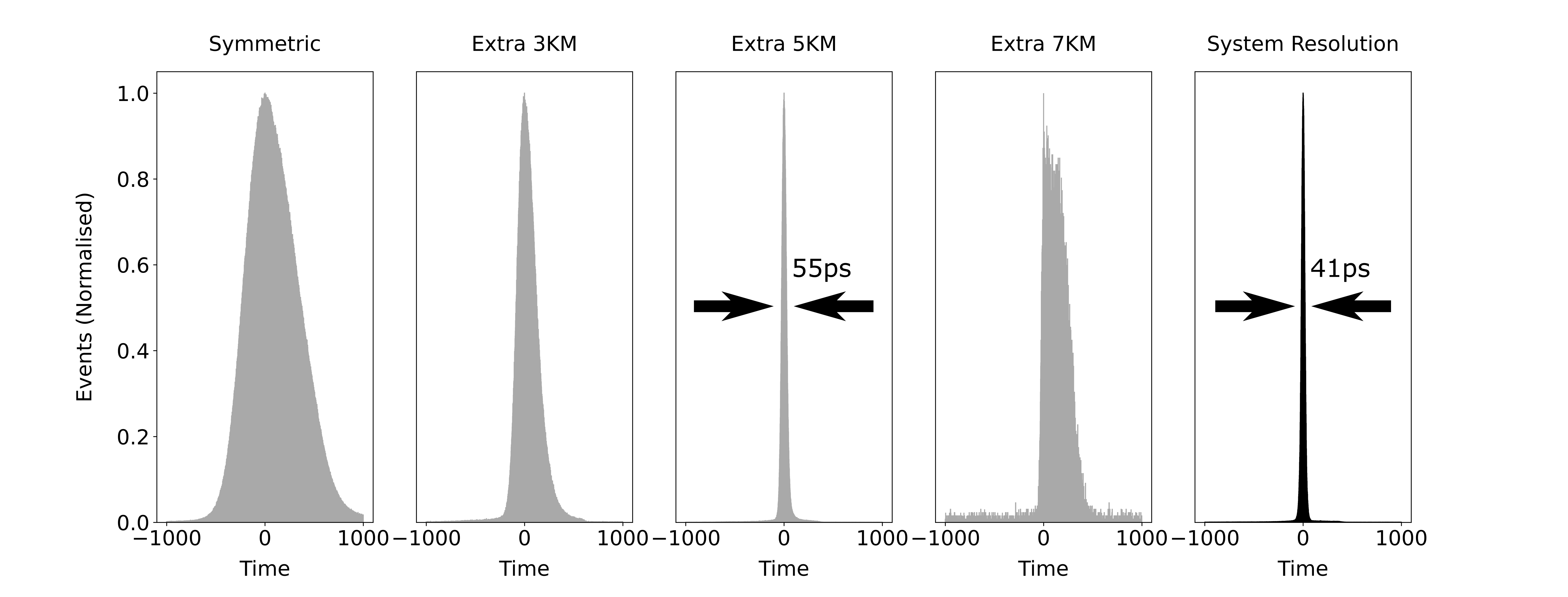}

\caption{From left to right) C($\tau$) plot when the signal and idler photons were propagated symmetrically (gray). When the signal photons were propagated over additional lengths of fiber corresponding to 3km, 5km, and 7km respectively (gray). System resolution corresponding to C($\tau$) plot when the signal and idler photons were sent directly into the detectors (black). Peaks have been centralized at $\tau$ = 0 for ease of comparison.}
\label{Figure 6}
\end{figure}

\begin{figure}[hbt!]
\captionsetup{width=.8\linewidth}
\centering\includegraphics[width=10.5cm]{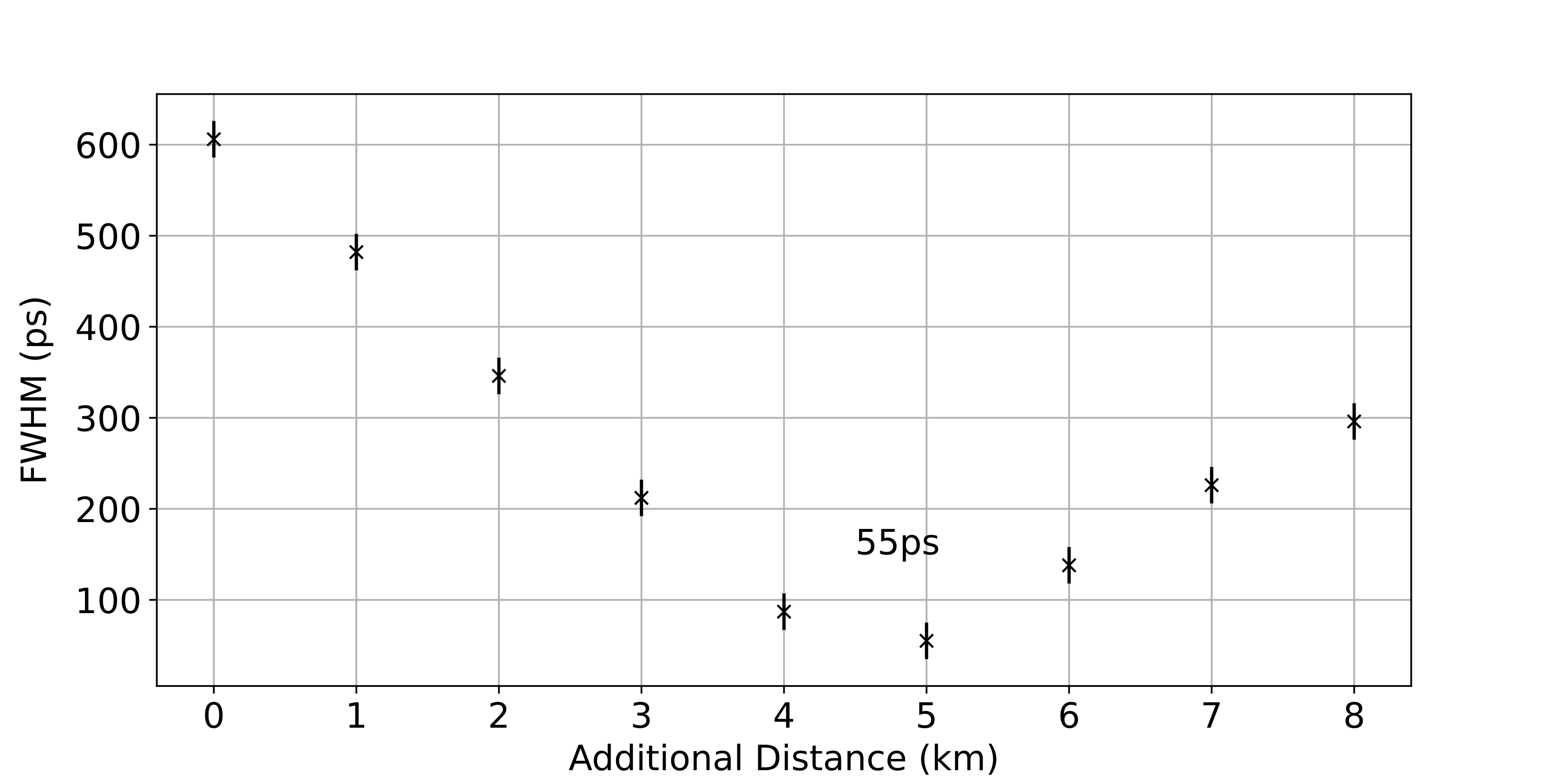}

\caption{Measured FWHM of the C($\tau$) plots as a function of the extra distance travelled by the signal photons after symmetric propagation over 10km of lab fiber.}
\label{Figure 7}
\end{figure}

\newpage 

\begin{figure}[hbt!]
\captionsetup{width=.8\linewidth}
\centering\includegraphics[width=9cm]{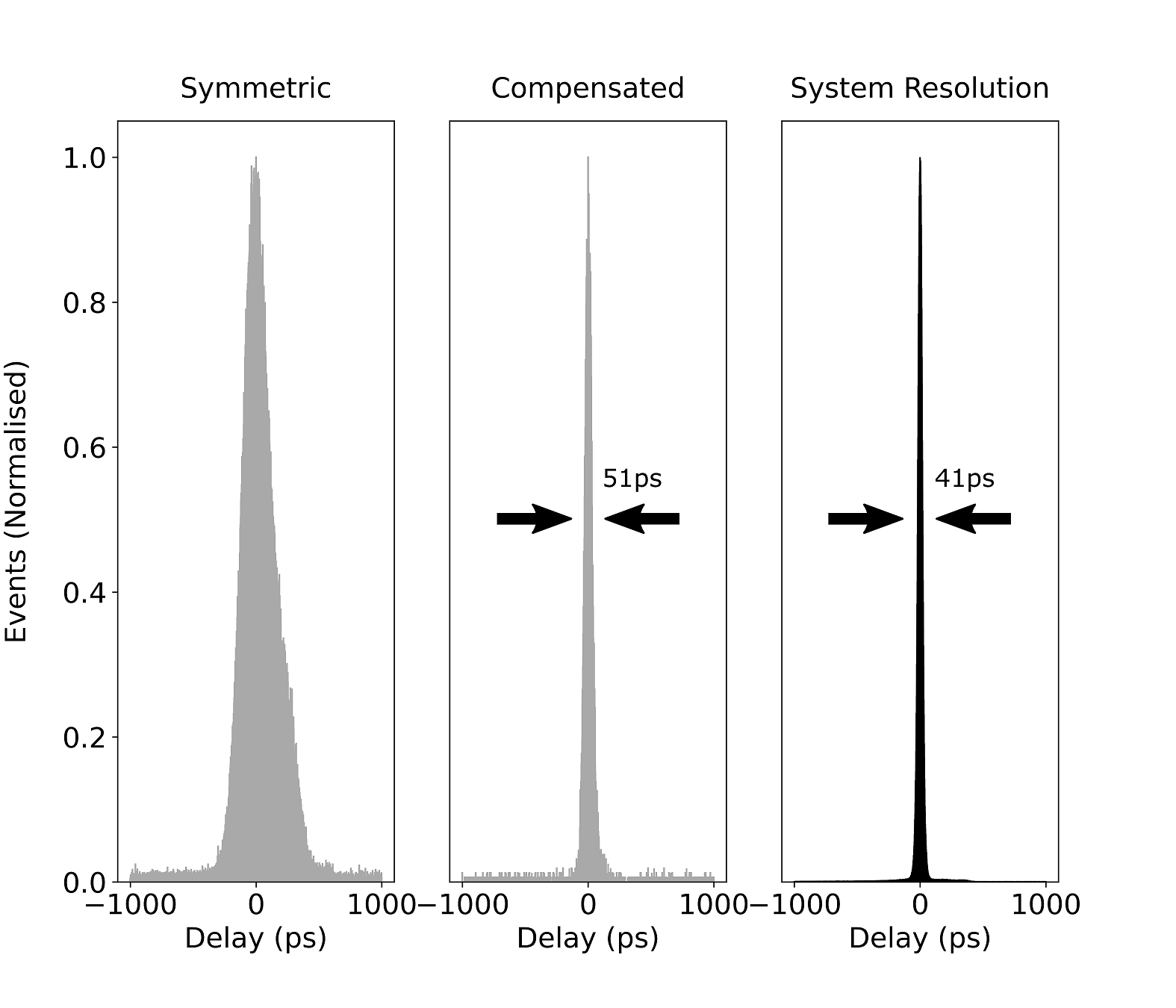}

\caption{Left to right) C($\tau$) plot when the signal and idler photons were propagated symmetrically over two multi-segmented 10km spans deployed fiber (gray). Near-ideal compensation was reached when the signal photons were propagated over additional lengths of fiber corresponding to 2km where the FWHM of timing correlations were preserved within 51ps (gray). System resolution corresponding to C($\tau$) plot when the signal and idler photons were sent directly into the detectors (black). Peaks have been centralized at $\tau$ = 0 for ease of comparison.\parfillskip0cm}
\label{Figure 8}
\end{figure}

Due to the difference in the magnitude of dispersion experienced by the signal and idler photons for the same propagation distance, there is still an increase in the width of the C($\tau$) peak. The degree of compensation can be maximized by propagating the less dispersed signal photons through additional lengths of fiber in order to introduce equal amounts of dispersion between the signal and idler photons. From the results, this correction could be achieved for our source with a precision on the order of 100-200ps/km, thus, 1km lengths should be suitable for achieving timing correlations within this value.\\

We thus experimentally investigated the effect of additional fiber in 1km lengths. The cross correlation was performed on correlated photon pairs that had both been propagated over the same piece of fiber 10km in length. Figure \ref{Figure 6} shows the evolution of C($\tau$) while Figure \ref{Figure 7} summarizes the results quantified in terms of the FWHM of the C($\tau$) peaks. Our results confirm that near-ideal nonlocal dispersion compensation was reached as evidenced by the preservation of the C($\tau$) peak’s width, with the measured value lying close to the limit imposed by our system jitter. Operationally, this procedure could easily be performed simply with the use of inexpensive compact fiber spools. Finally, using this technique we obtained timing correlations of 51ps for photon pairs propagated through separate 10km spans of multi-segmented deployed fiber. This result was obtained by propagating the signal photons with additional lengths of fiber corresponding to 2km (see Figure \ref{Figure 8}).\parfillskip0cm

\newpage

\section{Conclusion \& Discussions\raggedright}

The International Telecommunication Union specifies designated chromatic dispersion parameters for optical fibers in order to simplify the incorporation of chromatic dispersion compensation schemes into system design. This is in turn beneficial for our technique. The most simplistic scenario corresponds to the case where near-ideal compensation is reached with a fixed ratio of the propagation distances for the signal and idler photons of a particular source. However, the difference in the required length of fiber in the lab and field tests suggests that this standard may not be sufficiently precise to achieve this level of simplicity to the system design. Nevertheless, this limitation is not insurmountable since this technique does not require highly precise tuning to obtain near-ideal timing correlations. Furthermore, a one-time characterization of the fiber-optic network together with the compensating fiber spools can be used as a deterministic solution to the problem. Finally, the effect of variability in the dispersive properties of the fiber can be minimized by constraining the possible disparity in the chromatic dispersion parameters between the correlated photon pairs\textemdash if the SPDC photon pairs were produced about the median value of the range of possible zero dispersion wavelengths (1312nm).\break

Dispersion and nonlocal dispersion compensation of broadband photons pairs propagated over ITU-T G.652D telecommunication fiber were studied extensively in this paper. We show that tight timing correlations can be obtained by tuning the propagation distance in a practical manner. Telecommunication fiber-based nonlocal dispersion compensation between broadband photon pairs has the potential to supplement O-band quantum communication. Single mode fibers possessing zero dispersion wavelengths within the O-band are used extensively in relatively short-haul networks such as Metropolitan Area Networks (MAN), primarily for its low cost or the need for high density usage\cite{FS,ITU-TG657}. Quantum communication in the O-band may be able to share fiber with classical traffic in the C-band\cite{LiuBroadband}. We believe that this technique may find uses in scenarios relying on the coincident detection of photon pairs, for example, in increasing the bit-rate to reducing the error rate in quantum key distribution to improving the precision of SPDC-based methods of clock synchronization\cite{Lee:16,Xu2019Energy,Shi2020stable,Lee2019symmetrical}.\break

\section*{Funding}

This research is supported by the National Research Foundation, Prime Minister's Office, Singapore under its Corporate Laboratory@University Scheme, National University of Singapore, and Singapore Telecommunications Ltd.\\

\section*{Acknowledgements}

We would like to thank the Singtel fiber team for facilitating our deployed fiber tests.
\newpage


\bibliographystyle{unsrt}
\bibliography{Citations.bib}

\end{document}